\documentclass[nocompress]{spie}

\usepackage{amsmath,amsfonts,amssymb}
\usepackage{graphicx}
\usepackage[colorlinks=true, allcolors=blue]{hyperref}

\title{Filter and baffle design for the short-wavelength camera on SOLAR-C EUVST}

\author[a,b,*]{James McKevitt}
\author[a]{Duncan Rust}
\author[a]{Berend Winter}
\author[a]{David Walton}
\author[a]{Sarah Matthews}
\author[a]{Ted Pyne}
\author[c]{Yukio Katsukawa}
\author[c]{Ryohko Ishikawa}
\author[d]{Alexey Shitvov}
\author[a]{Chloe Wishart}
\author[a]{Barry Whiteside}
\author[e]{Jacob Parker}
\author[a]{Diagarajen Carpanen}
\affil[a]{University College London, Mullard Space Science Laboratory Holmbury St Mary, Dorking Surrey, RH5~6NT, UK}
\affil[b]{University of Vienna, Institute of Astrophysics Türkenschanzstrasse 17 Vienna A-1180, Austria}
\affil[c]{National Astoronomical Observatory of Japan, 2-21-1 Osawa, Mitaka, Tokyo 181-8588, Japan}
\affil[d]{University College London, Department of Physics and Astronomy Gower St, London, WC1E~6BT, UK}
\affil[e]{Montana State University, Culbertson Hall, 100, Bozeman, MT~59717, United States}

\authorinfo{Further author information: (Send correspondence to J.M.)\\J.M.: E-mail: james.mckevitt.21@ucl.ac.uk}

\begin{document} 
\maketitle

\begin{abstract}
SOLAR-C, which is scheduled to launch in the late 2020s with the Extreme Ultraviolet High-Throughput Spectroscopic Telescope (EUVST), will deliver EUV spectroscopy of the solar atmosphere with unprecedented temperature coverage and resolution. We present the optical design and mechanical impacts for the short-wavelength camera, and how our design choices affect scientific performance using complete forward modelling. We discuss the design and optical impact of the thin-film aluminium filter and supporting mesh, including diffraction and shadowing effects in spectral and spatial dimensions. We analyse this using our own Fourier optics method and analytical modelling, and quantify the scientific impact using forward modelling driven by a simulated solar atmosphere. We address the vulnerability of thin-film filters during launch depressurisation using vacuum chamber testing. We describe the stray-light baffle design and low-scatter coatings which suppress visible stray light.
\end{abstract}

\keywords{Thin-film optical filter, Stray light, Extreme ultraviolet (EUV), Venting, Mesh diffraction, Forward modelling, Solar spectroscopy, Slit-scan spectrometer}

\section{INTRODUCTION}


The JAXA SOLAR-C mission, scheduled for launch in 2028, will carry the next-generation Extreme-Ultraviolet (EUV) Spectroscopic Telescope (EUVST), a slit-scan spectrometer integrated with a slit-jaw imager\cite{shimizu_solar-c_euvst_2019}. A solar slit-scan spectrometer isolates a narrow spatial slice of the solar disk via an entrance slit, passing this light to a diffraction grating that disperses the spectrum across a two-dimensional detector array. Light not passing through the slit is reflected to the context slit-jaw imager. In this way, spatially and spectrally-resolved information along one axis is captured at each slit position, and the orthogonal spatial dimension is sequentially constructed by stepping the slit across the solar disk. This fine spectral resolution of the plasma emission can then be used to determine plasma velocity, electron density, and composition, and other plasma parameters\cite{del_zanna_solar_2018}.

EUVST will provide an unprecedented ability to diagnose solar atmospheric plasma, with a spatial resolution down to $\sim$0.4~arcsec across a wide quasi-continuous temperature range of 0.02--15~MK ($\Delta\log T \leq 0.4$) at cadences down to 0.5~seconds. EUVST will observe the fundamental processes powering coronal heating, trace the pre-flare mechanisms that trigger solar eruptions, and resolve the physical consequences of magnetic reconnection in the solar atmosphere\cite{mckevitt_pre-flare_2026}.

\begin{figure*}
    \centering
    \includegraphics[width=\linewidth]{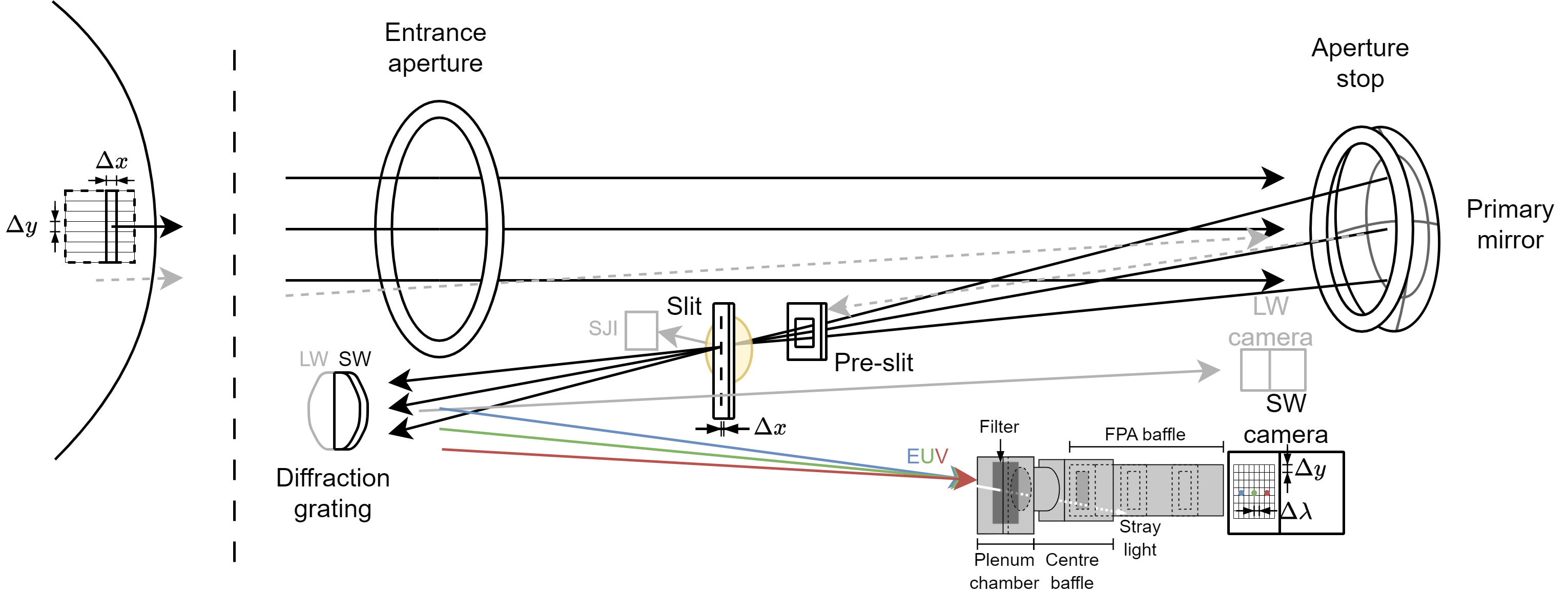}
    \caption{Schematic diagram of the SOLAR-C/EUVST short-wavelength (SW) channel (red/green/blue)\cite{mckevitt_solar-ceuvst_2026}, with the SW camera baffle detail added. Light enters through the entrance aperture and is focused by the primary mirror. Light from outside the field of view is practically fully absorbed by the pre-slit (dashed grey). The slit selects the spatial region to be observed, and the diffraction grating disperses the light spectrally. The thin-film aluminium filter blocks visible stray light (solid grey) while transmitting EUV radiation. We also show the slit-jaw imager (SJI) and the long-wavelength (LW) camera\cite{mckevitt_solar-ceuvst_2026}. The various SW baffle sections are explained in the text, where FPA baffle is the focal plane assembly baffle.}
    \label{fig:euvst_cartoon}
\end{figure*}

EUVST consists of a primary mirror (with an aperture stop diameter of 280~mm), various selectable slits (0.2, 0.4, 0.8, and 1.6~arcsec), and a split diffraction grating. This grating disperses a short-wavelength (SW; 17.0--21.2~nm) channel to one dedicated SW camera, and long-wavelength (LW; various bands between 46.4 and 122.1~nm) channels to a separate long-wavelength camera. The SW camera (EUVST-SW) is being built at UCL-MSSL for the European Space Agency, and the LW camera (EUVST-LW) is being built for NASA by the US Naval Research Laboratory (NRL). We show a schematic layout of the instrument in Figure~\ref{fig:euvst_cartoon}.

The physical processes which will be observed by EUVST occur on small spatial scales with subtle intensity variations and velocity signatures. Suppressing stray light and mitigating optical artefacts is, therefore, critical for the mission to meet its science requirements. In this paper, we describe the baffle structure and thin-film aluminium filter used to suppress stray light at the SW camera, the impact of the filter support structure's diffraction artefacts at the focal plane, our tests of the pressure differential we expect across the filter during spacecraft venting on launch, and the expected science performance as a function of filter parameters using the complete forward-modelling framework ECLIPSE.




\section{SW Camera Baffle Design and Coating}


Solar spectral irradiance peaks within the visible wavelength range (approximately 400--700~nm), and so a vast proportion of the solar radiation incident on the telescope aperture is out-of-band signal that we want to avoid reaching the focal plane. The previous generation EIS\cite{culhane_euv_2007} instrument on spacecraft Solar-B\cite{kosugi_hinode_2007} (renamed Hinode after launch) incorporated a visible-light filter at the telescope entrance to reject all but $\sim$$10^{-8}$ of incident visible light\cite{korendyke_optics_2006}, which practically precluded visible light entering the telescope. However, SOLAR-C covers a significantly broader EUV wavelength range than Hinode, and no single entrance filter is capable of transmitting this wide EUV bandwidth while simultaneously blocking visible light.

EUVST is designed to a strict visible stray-light requirement of 1~photon/pixel/second incident on the detectors\cite{tsuzuki_high-fidelity_2026}, a threshold beyond which additional stray light degrades the spectrograph's ability to accurately measure small intensity and velocity variations in the plasma. Such issues with stray light in EUV spectrograph measurements can be seen explored using data from Hinode/EIS\cite{young_scattered_2022}. The microchannel plate (MCP) intensifiers used in the LW channels use coatings with photoelectric thresholds exceeding visible-light photon energies, and so carry an effective visible-light rejection estimated to be on the order of $\sim$10$^{-5}$. In the short-wavelength camera however, the Te2v CCD42-40 NIMO detectors detect visible-light photons with a quantum efficiency (QE) approaching 50\%\cite{e2v_ccd42-40_2016}. Therefore, visible light must be eliminated before it reaches the SW focal plane. To achieve this, the opto-mechanical design of the instrument relies on high-absorptivity coatings on critical internal surfaces, internal mechanical structures designed to intercept first- and second-order scatter paths, and an optical filter mounted ahead of the focal plane.


\subsection{Baffle design and vane positioning}

\begin{figure*}
    \centering
    \includegraphics[width=\linewidth]{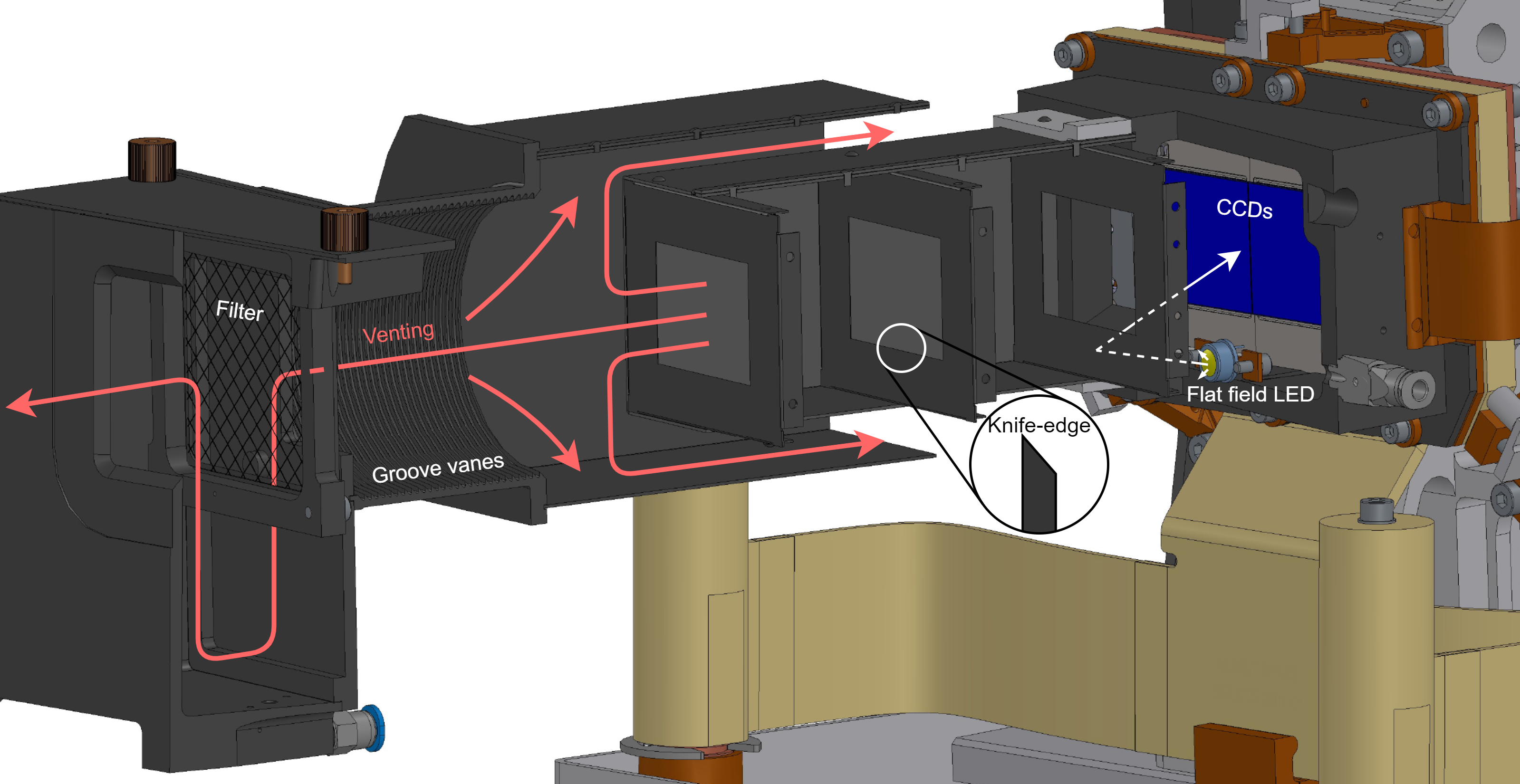}
    \caption{Schematic of the SW camera baffles, showing the optical filter on entrance to the baffle structure, internal vanes with knife-edges, and an LED installed to provide diffuse quasi-uniform illumination of the CCDs to help with in-flight calibration. Annotated are the pathways for air to vent during launch. Note the filter mesh shown here is not to scale.}
    \label{fig:sw_baffle}
\end{figure*}


The SW camera baffle is designed to accept the SW channel reflected from the diffraction grating while rejecting off-axis stray light (shown in Figure~\ref{fig:euvst_cartoon}). The baffle is formed of two halves, seen in Figure~\ref{fig:sw_baffle}. The first half, fixed to a spacecraft bulkhead, houses the optical filter in a `plenum chamber', which then connects via a cylindrical section with groove vanes to a wider box section. We term this half the `centre baffle'. The second half, attached to the SW camera and extending into the centre baffle, consists of a rectangular box section with three internal vanes. It is mounted to the camera via a radiation shield which surrounds the CCDs and two internal light-emitting diodes (LEDs). We term this baffle the `FPA baffle', where FPA stands for focal plane assembly. The payload fairing, spacecraft, and instrument components will launch at ambient pressure and vent to vacuum during launch, so we also annotate the venting paths of this baffle structure.

The baffle section is constructed of two halves to allow movement of the camera during its pre-flight alignment to ensure the CCDs are perfectly centred at the focal plane of the optical system, but still allow the baffle structure to remain light-tight to its fixed mounting at the front bulkhead. As such, clearance between the FPA and centre baffles is specified according to the instrument alignment budget, where this gap also makes a venting path available. The plenum chamber in which the optical filter is mounted is designed to reduce the air speed around the filter during the launch venting, and so also the stress on the filter. This diffusion effect requires the feed volume to be narrower in cross-section than the plenum chamber. As the cylindrical volume cannot therefore be expanded, full vanes to avoid grazing incidence reflections are not possible without vignetting the beam. We therefore use grooved vanes in the cylindrical walls to suppress grazing-incidence scatter\cite{fest_stray_2013}.

Three internal vanes in the FPA baffle reject first-order stray light paths which reflect off the internal surfaces of the main baffle, while not vignetting the SW beam\cite{fest_stray_2013}. The final vane is also positioned to allow diffuse reflection of light from two light-emitting diodes (LEDs) mounted inside the radiation shield to illuminate the detector. These can be turned on when the shutter is closed to make flat-field measurements, a process which is used to help track detector contamination over the mission lifetime, as has been done on Hinode/EIS\cite{bradley_contamination_2011}. These vanes all feature knife-edges to minimise the surface area at the vane aperture boundaries available for grazing incidence scattering.

\subsection{High-absorptivity coating}


Plasma Electrolytic Oxidation (PEO) is an electrochemical surface treatment that converts the surface layers of light metals, aluminium in this case, into a hard ceramic oxide. Through the plasma-assisted growth process, optically dark species can be incorporated into the ceramic structure, producing a highly absorptive black finish \cite{shrestha_preliminary_2003}. Such a coating, applied by Keronite UK, is used on the SW camera and baffle structure to reduce stray light by suppressing specular reflections and absorbing most incident optical radiation. The measured solar absorptance of the coating, $\alpha = 0.94$, corresponds to a total solar reflectance of approximately 0.06. In our coating qualification campaign, performed in accordance with ECSS-Q-ST-70-17C, the thermo-optical properties $(\alpha = 0.94, \epsilon = 0.83)$ were found to be stable against solvent cleaning, light abrasion, humidity exposure, and vacuum thermal cycling. The use of this optically black coating therefore introduces a thermal-control trade-off: while beneficial for stray-light suppression, its high solar absorptance increases the absorption of radiative parasitic heat loads by the baffle. This is important because the CCD must be maintained close to its target operating temperature of $\lesssim -60^\circ\mathrm{C}$, and is thermally coupled to the baffle and radiation-shield assembly primarily through conductive paths, with a smaller radiative coupling also present. Thermal analysis shows that the resulting penalty is acceptable given the corresponding reduction in stray light within the telescope.

\section{Thin-film aluminium filter design and venting}

The optical filter used to reject visible light from the SW baffle structure is a 150~nm thick thin-film aluminium filter manufactured by Luxel. This thickness is possible to manufacture, has a high visible light rejection rate, and still has a relatively high EUV transmission rate. However, unsupported aluminium of this thickness cannot survive standard acoustic launch loads, so the film is bonded to a supporting nickel mesh with 390~$\mu$m pitch and a 40~$\mu$m bar width, as used on Hinode/EIS\cite{korendyke_optics_2006}. Similarly, we also implement stress-reducing filleted corners at the mesh-frame interface, found to be essential for the survivability of the filter during pre-flight testing for Hinode/EIS. We also orient the mesh at 45\textdegree to the CCD pixel arrangement to prevent diffraction peaks generated by the mesh from aligning with the sensor's read-out axes to mitigate blooming. The spacecraft and internal components will launch without a vacuum housing, and so venting of the internal structure will occur during the launch. In this section, we describe our efforts to estimate the pressure differential experienced by the filter and the optical impact of the filter and supporting mesh on the SW beam.


\subsection{Optical performance}

The 150~nm filter is expected to contain $\sim$1\% embedded aluminium oxide. A further passivation layer of aluminium oxide is expected to form on each side of the filter aluminium during its storage in air on the ground before launch. This can be mitigated using vacuum storage\cite{powell_care_1993}, but given the risk posed to the fragile filters every time they are removed and re-stored during periodic pre-flight testing, we instead use a high-purity dry nitrogen purge to minimise the humidity of their storage environment and slow the formation of this layer. This purge can also help avoid the formation of a second layer of molecular carbon contamination, which is anticipated to accumulate on the filters over time. The aluminium oxide layer is expected to saturate at approximately 4~nm on each side of the filter, and the contamination control plan for the filters is designed to limit carbon contamination to 4~nm on each side of the filter.

\begin{figure}
    \centering
    \includegraphics[width=0.5\linewidth]{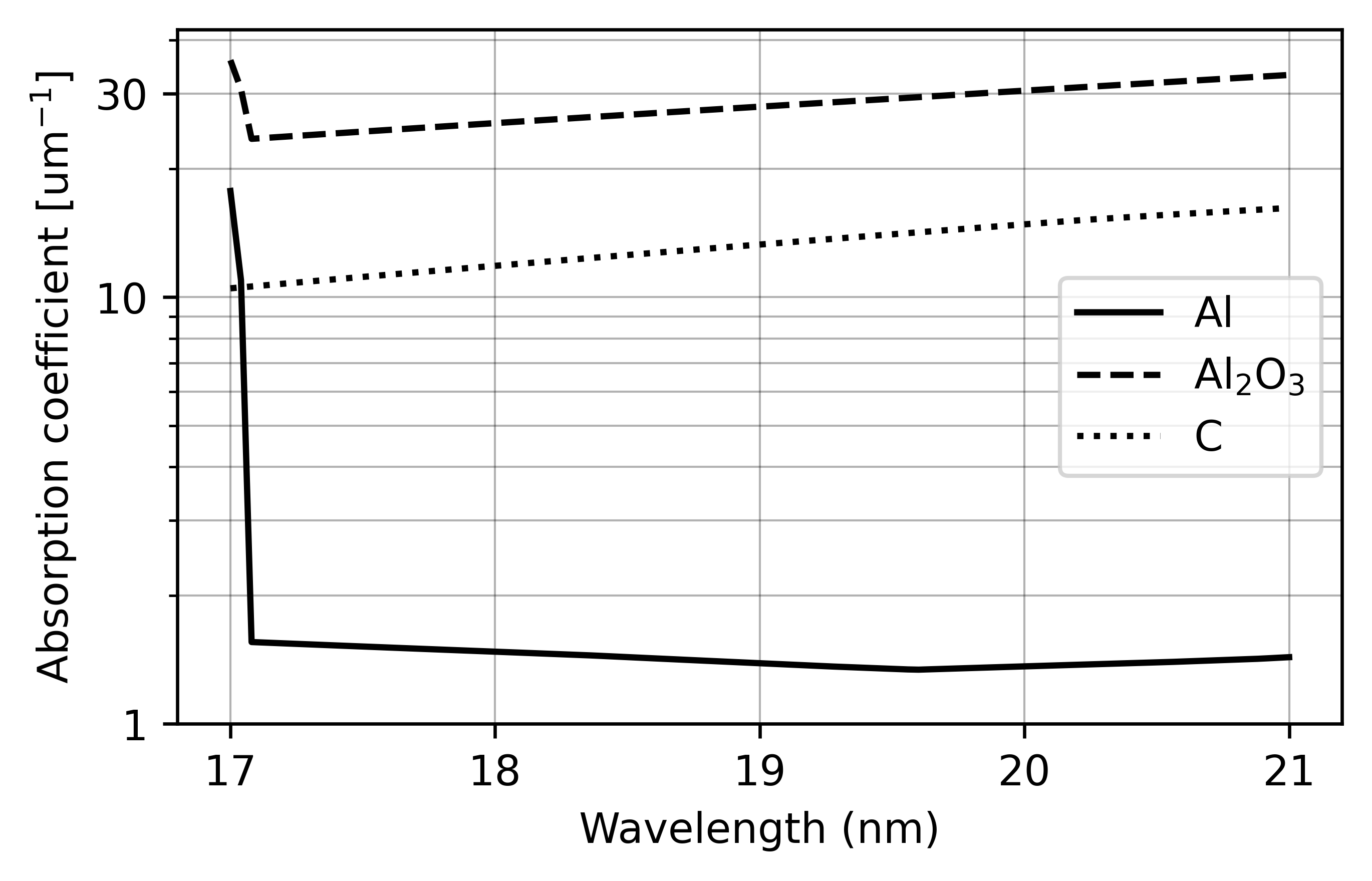}
    \caption{Absorption coefficients of aluminium (Al), aluminium oxide (Al$_2$O$_3$), and carbon (C)\cite{henke_x-ray_1993}.}
    \label{fig:absorb_coefficients}
\end{figure}

The absorption coefficient, $\alpha$, quantifies the rate at which electromagnetic radiation attenuates as it propagates through a medium. It relates to fractional transmission, $T$, via the Bouguer-Lambert law, $T = \exp(-\alpha x)$, where $x$ is the material thickness\cite{bouguer_essai_1729,lambert_photometria_1760}. Comparing the coefficients for aluminium, aluminium oxide, and carbon, we can calculate the anticipated EUV throughput of the composite filter stack. We use the values shown in Figure~\ref{fig:absorb_coefficients} for this, and find them in agreement with empirical results\cite{powell_thin_1990}. Using the anticipated aluminium, aluminium oxide, and carbon thicknesses (148.5~nm, 9.5~nm, 8~nm respectively), and the mesh throughput (80\%), we expect the total filter transmission in the SW channel to be $\sim$50\% at 19.5~nm.

In the visible wavelength range, the filter is expected to provide a throughput on the order of $10^{-7}$\cite{korendyke_optics_2006}. Pinholes on the scale of 1~$\mu$m, inherent to the manufacturing process, introduce minor localised variations in this transmission factor.



\subsection{Filter pressure differential and venting tests}

\begin{figure}
    \centering
    \includegraphics[width=0.45\linewidth]{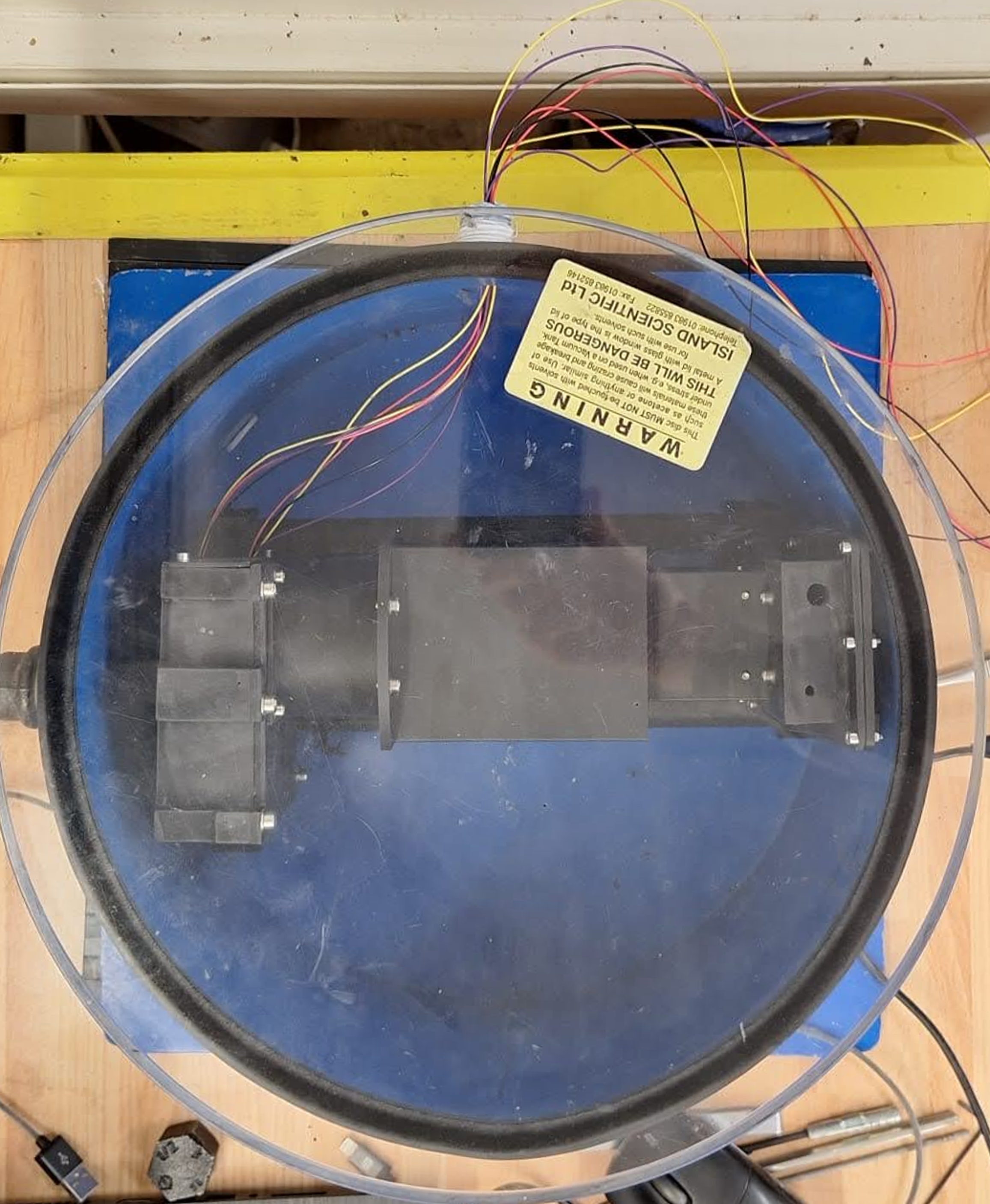}
    \caption{Test setup showing 3D-printed baffle structure inside vacuum chamber where wires are connected to two pressure transducers on either side of a dummy filter.}
    \label{fig:filter_pressure_test}
\end{figure}

\begin{figure*}
    \centering
    \includegraphics[width=0.75\linewidth]{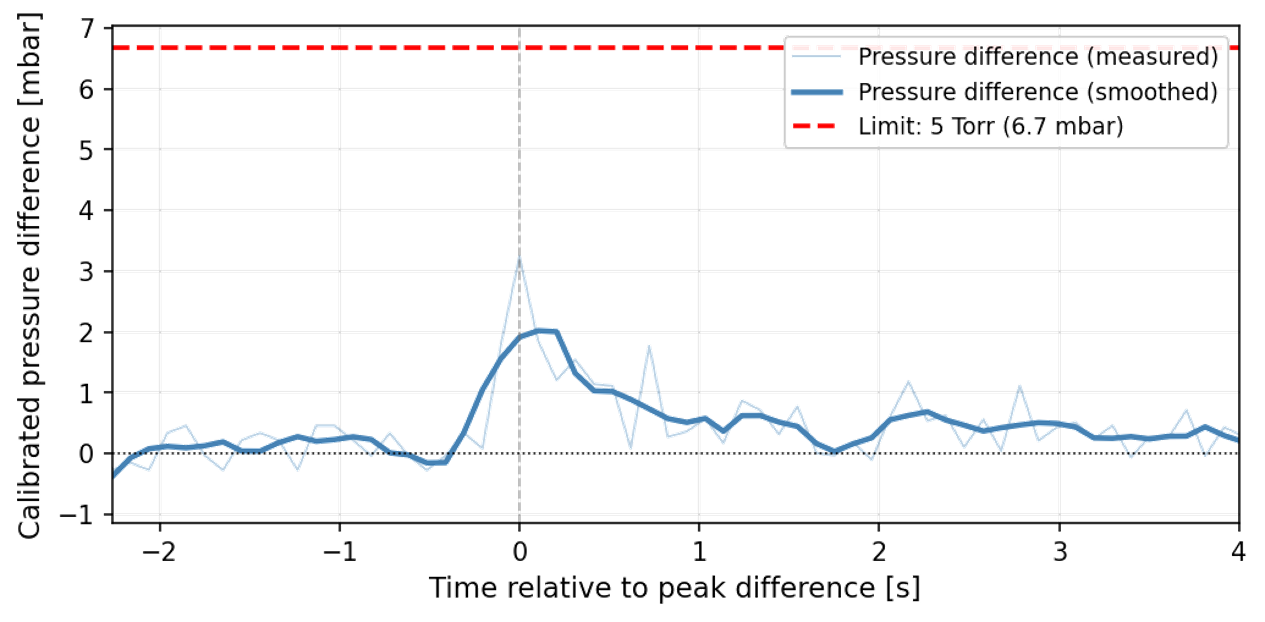}
    \caption{Pressure differential measured across the thin-film aluminium filter during vacuum chamber testing, where the chamber was evacuated from ambient to vacuum in $\sim$60~s to conservatively approximate a typical launch vehicle profile\cite{arianespace_ariane_2016}. The peak difference here corresponds to the fastest rate of pressure decrease as approximated from the launch profile.}
    \label{fig:filter_pressure_diff}
\end{figure*}


As the filter will not be held in a vacuum enclosure during the launch, it must survive the ambient pressure decay inside the rocket payload fairing, and the associated pressure differential across the filter as the baffle vents (Figure~\ref{fig:sw_baffle}). To evaluate this expected differential, we 3D printed a mock-up baffle structure and installed a dummy filter sandwiched between two pressure transducers. We put this assembly inside a vacuum chamber and manually depressurised the system at a rate approximately equal to that expected in the payload fairing, using the Ariane~5 profile as a guide\cite{arianespace_ariane_2016}. We conservatively took this time to be 60~s to hard vacuum to account for the true non-linear profile we could not accurately replicate with manual valve control. The optical filter is a critical single point of failure for the instrument channel, meaning we took a conservative maximum pressure differential limit of 5~Torr ($\sim$6.7~mbar). The peak pressure differential measured during the rapid chamber evacuation remained below this threshold, as shown in Figure~\ref{fig:filter_pressure_diff} around the peak time of differential pressure.



\section{Mesh Diffraction}

The nickel mesh needed to help the filter withstand the aforementioned pressure differentials and acoustic launch loads introduces an opaque periodic spatial filter into the optical path, which causes a complex diffraction pattern at the image plane. Artificial contrasts and secondary diffraction peaks generated by such structures frequently complicate the analysis of high-resolution solar data, particularly during high-intensity events\cite{lin_diffraction_2001}.

\subsection{Mesh diffraction modelling}

Modelling the diffraction caused by the mesh is complex in our case due to the converging nature of the beam. We approximate this problem using a converging beam of uniform intensity at 19.5~nm focused at an image plane a distance $z_2$ from the mesh. The beam is approximately 0.25~cm in radius at the mesh and passes perpendicularly through the mesh as it propagates 25~cm to the detector. This geometry is a reasonable approximation of the real optical design, made using Ansys Zemax OpticStudio.

\subsubsection{Numerical Fourier optics}

To evaluate the mesh diffraction we implement a custom numerical model based on the angular spectrum method, using the Fourier optics principle that any complex optical field can be decomposed into a continuous superposition of uncoupled plane waves travelling at distinct angles\cite{goodman_introduction_2017}.

Because the beam converges towards a focus at a distance $z_2$ its initial incident wavefront is curved. We define the initial complex electric field immediately preceding the mesh plane, $U(x,y,0)$, by applying a spherical phase factor, $\phi(R)$, to a uniform scalar amplitude $A_0$. This phase factor accounts for the geometric path difference of rays converging from the filter plane to the focal point such that

\begin{equation}
    \phi(R)=\exp\left[-ik\left(\sqrt{R^2+z_2^2}-z_2\right)\right]
\end{equation}

\noindent{}where $R=\sqrt{x^2+y^2}$ is the radial distance from the optical axis across the beam cross-section, and $k=2\pi/\lambda$ is the wave number. The unperturbed incident field then takes the form

\begin{equation}
    U(x,y,0)=A_0\cdot\phi(R).
\end{equation}

The nickel mesh can be considered a binary spatial filter, opaque at the bars and transparent at the open apertures. Multiplying the incident field by the binary transmission function of the mesh, $T_\text{mesh}(x,y)$ can be used to find the field immediately following the mesh

\begin{equation}
    U_\text{mesh}(x,y)=U(x,y,0)\cdot T_\text{mesh}(x,y).
\end{equation}

This field is then propagated over the free-space distance $z_2$ to the image plane. To do this, we use a 2D spatial Fourier transform, which decomposes the complex beam to its angular spectrum $A(f_x,f_y)$, where $f_x$ and $f_y$ are the spatial frequencies. We advance each independent plane wave over the distance $z_2$ by applying a free-space transfer function

\begin{equation}
    H(f_x,f_y)=\exp\left(i2\pi z_2\sqrt{\frac{1}{\lambda^2}-f_x^2-f_y^2}\right).
\end{equation}

An inverse Fourier transform of this propagated spectrum results in the final complex field at the image plane, $U_\text{image}(x,y)$. The observable photon intensity profile can then be calculated using

\begin{equation}
    I(x,y)=\left|U_\text{image}(x,y)\right|^2.
\end{equation}

\begin{figure}
    \centering
    \includegraphics[width=0.5\linewidth]{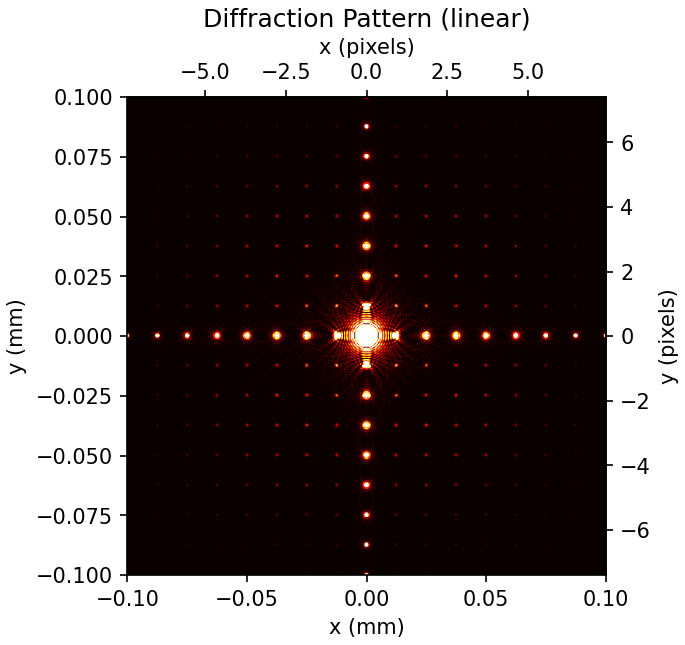}
    \caption{The diffraction pattern expected at the short-wavelength channel focal plane as calculated using Fourier optics, with a linear colour table. We show the spatial axes in physical and CCD pixel units where the EUVST-SW CCDs are 13.5x13.5~$\mu$m$^2$.}
    \label{fig:diffraction_pattern}
\end{figure}

We show the expected diffraction pattern generated using this method in Figure~\ref{fig:diffraction_pattern}.

\subsubsection{Analytical comparison}

We compared our 2D numerical simulation against a 1D analytical grating model\cite{born_principles_1999}. Because the mesh is a periodic grid we can approximate its behaviour as a standard 2D diffraction grating. In this idealised case, classical optics can be used to find the locations and intensities where diffracted plane waves will constructively interfere. The location of the $m$-th order diffraction peak at the image plane, $x_m$, can be found using the grating equation

\begin{equation}
    x_m=\frac{m\lambda z_2}{p}
\end{equation}

\noindent{}where $p$ is the grid pitch.

The peak intensity of each diffracted order relative to the central zero-order maximum ($I_0$) scales according to the diffraction envelope, defined by the individual rectangular apertures. Expanding the square-wave transmission function as a Fourier series results in the intensity ratio

\begin{equation}
    \frac{I_m}{I_0}=\text{sinc}^2\left(\frac{mw}{p}\right)
\end{equation}

\noindent{}where $w$ is the physical width of the open aperture in the mesh.

\begin{figure}
    \centering
    \includegraphics[width=.85\linewidth]{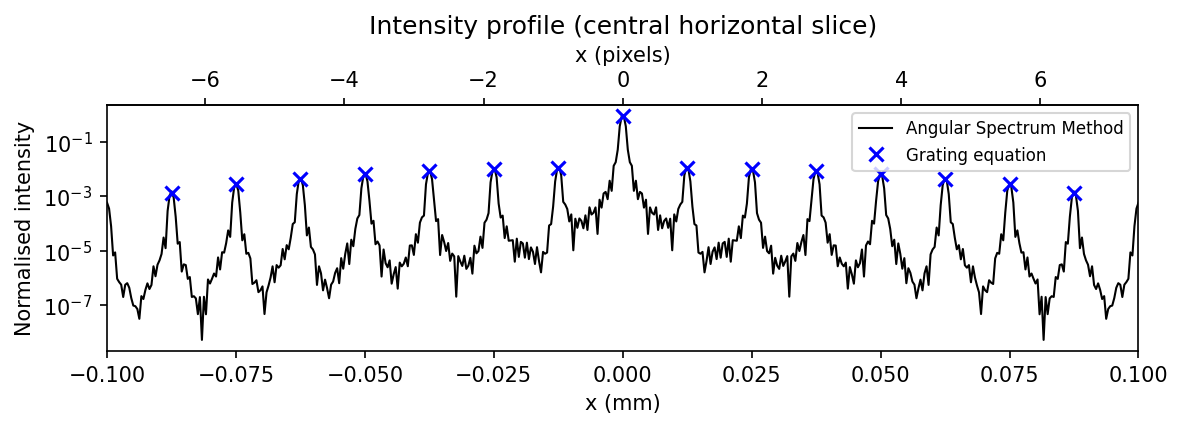}
    \caption{The diffraction pattern expected at the short-wavelength channel focal plane as calculated using Fourier optics and classical optics.}
    \label{fig:diffraction_profile}
\end{figure}

The result of both of these methods are shown together in Figure~\ref{fig:diffraction_profile}, where the intensity from the Fourier optics approach is shown at the central horizontal slice. We find them both to be in agreement, and that the $m>1$ order diffraction peaks have on the order 1\% the intensity of the central peak. Such a low level of contrast does not compromise the science performance of the instrument. Extreme cases where reconnection-induced high-intensity emission during a solar flare is captured against a dark background, such as off-limb observations, will be explored in future work.

\section{Science Impact and Forward Modelling}

\begin{figure*}
    \centering
    \includegraphics[width=\linewidth]{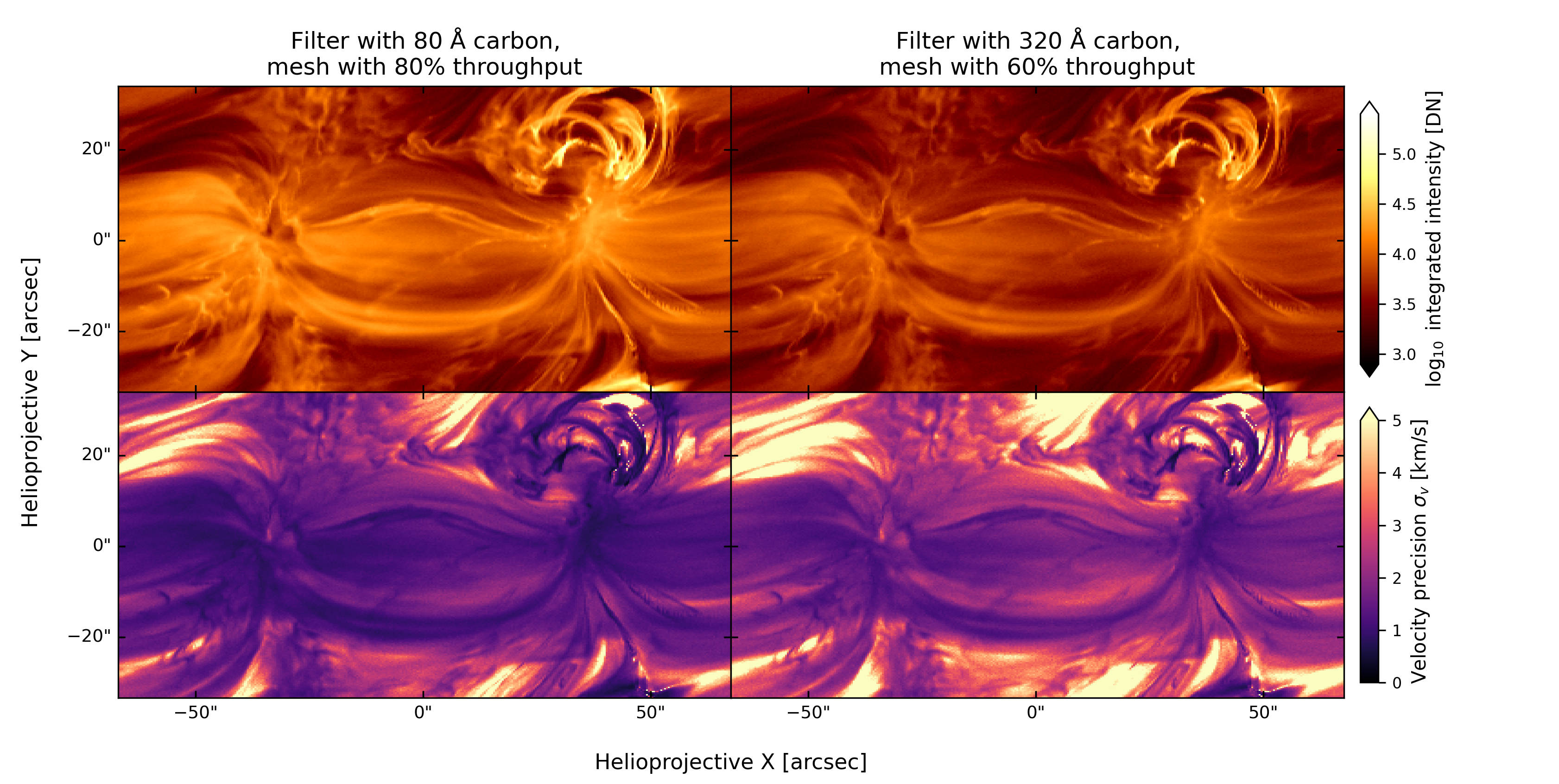}
    \caption{A comparison of measured intensity (top row) and the precision of plasma velocity measurements (bottom row) using the Fe~XII~195.119~\AA{} (log~T[K]$\sim$6.2) emission line for a fixed exposure time of 40~s per slit position, for different levels of filter carbon contamination and mesh throughout. The simulated observation uses the instrument's 0.4~arcsec slit and 2-pixel along-slit binning.}
    \label{fig:forward_model}
\end{figure*}

To consider the science impact of the thin-film aluminium filter and the various configurations it could take, we used our complete forward model of the instrument codified in the ECLIPSE code\footnote{\url{https://github.com/jamesmckevitt/eclipse}}. This allows us to quantitatively consider the scientific impact of specific engineering parameters, and directly trace them to our ability to meet the instrument's science requirements. The ECLIPSE code\cite{mckevitt_pre-flare_2026} synthesises spatially and spectrally-resolved optically-thin emission from a magnetohydrodynamic (MHD) simulation of the solar atmosphere, then forward models the instrument response to this atmosphere to generate a hyper-realistic simulated observation considering all relevant noise sources, throughput impacts, and optical artefacts. In this case, we simulate the instrument's response to a pre-computed MHD simulation of a solar active region $\sim$30~minutes before a flare\cite{cheung_comprehensive_2018}, where the instrument measured the Fe~XII~195.119~\AA{} emission line (formed at log~T[K]$\sim$6.2) in a spectral window containing the blended Fe~XII~195.179~\AA{} and the 10 brightest background emission lines.

We demonstrate the forward modelling code in Figure~\ref{fig:forward_model} by showing the performance of the instrument for different thicknesses of carbon contamination and mesh throughputs, where the left column is the filter configuration being used in the mission and the right column is a case with more structural support for the filter from the mesh but lower optical throughput, and higher carbon contamination. Both cases use 1\% embedded and 8~nm of surface (saturated) aluminium oxide (9.5~nm total). The top row of panels shows the intensity of emission measured in DN, integrated along the spectral direction of the detector. The bottom row shows the precision of our measurement of the plasma velocity, where this is calculated using a monte carlo fitting method to spectra convolved with shot noise, Fano detection noise, and instrument electronic noise. Lower intensity pixels have a lower signal-to-noise ratio, and so velocity is measured less precisely in lower-intensity regions.

Such a complete forward modelling tool is powerful in that it shows us directly the scientific impact of a different filter configuration and less stringent contamination control plan. We consider this against the instrument requirement to measure plasma velocity with a precision of 2~km/s. In the expected filter configuration (left), the plasma across most of the active region is measured to 2~km/s precision, while in this less optimal configuration (right), the plasma is mostly measured to 4~km/s precision. For a standard exposure time of $\sim$40~s, this allows us to determine that the nominal filter configuration and contamination level are sufficient to meet the science requirement of 2~km/s Doppler velocity precision, while any decrease in mesh throughput and increase in carbon contamination would challenge our ability to meet this requirement.



\section{Conclusions}

The short-wavelength filter and baffle assembly for SOLAR-C/EUVST represents a design capable of meeting competing thermal, mechanical, and optical requirements. To prevent visible stray light from reaching the EUVST-SW focal plane, a thin-film aluminium filter is required. Although this practically removes all visible light, its intrinsic EUV throughput coupled with unavoidable oxidisation and contamination impact the science performance of the instrument. Furthermore, given its fragility it requires a supporting mesh to enhance its structural integrity during launch. We performed a test of the anticipated pressure differential across the filter during launch and found it to be acceptable. This supporting mesh provides a blocking factor which further impacts EUV throughput and introduces unavoidable diffraction effects at the focal plane during operation. We modelled the anticipated mesh diffraction pattern and intensity, and found it to be negligible for the mission's primary science, and performed a complete forward model of an observation by the instrument and found the nominal filter and mesh configuration to allow the instrument to meet the mission science requirements.

\acknowledgments
 
J.M. was supported by STFC PhD Studentship number ST/X508858/1. J.M., D.R., B.W., D.W., S.M., T.P. and D.C. acknowledge support from ESA Contract No. 4000141160/23/NL/IB.

\bibliography{references}
\bibliographystyle{spiebib}

\end{document}